\documentclass[aps,superscriptaddress,amsmath,twocolumn,amssymb]{revtex4}
\usepackage{epsfig}

\begin{document}

\title{Decoupling of self-diffusion and structural relaxation during a fragile-to-strong crossover in a kinetically constrained lattice gas}

\author{Albert C. Pan}

\affiliation{Department of Chemistry, University of California,
Berkeley, CA 94720-1460}

\author{Juan P. Garrahan}

\affiliation{School of Physics and Astronomy, University of
Nottingham, Nottingham, NG7 2RD, UK}
 
\author{David Chandler}

\affiliation{Department of Chemistry, University of California,
Berkeley, CA 94720-1460}

\date{\today}

\begin{abstract}
We present an interpolated kinetically constrained lattice gas model
 which exhibits a transition from fragile to strong supercooled liquid
 behavior.  We find non-monotonic decoupling that is due to this
 crossover and is seen in experiment.
\end{abstract}

\maketitle

\twocolumngrid

The viscosity of glassformers increases by several orders of magnitude
upon cooling near the glass transition
temperature.  
For strong glassformers,
this increase is Arrhenius, that is, the viscosity increases exponentially with inverse
temperature, whereas, for fragile glassformers, this increase is
super-Arrhenius \cite{Ediger-et-al,Angell,Debenedetti-Stillinger}.  
Recent experimental and theoretical evidence indicate that
fragile glasses can undergo a crossover to strong behavior at low enough
temperatures \cite{Ito-et-al, Buhot-Garrahan, Garrahan-Chandler}.  
Here, we show that a fragile-to-strong
crossover can be modeled by a union of two kinetically constrained
lattice gas models \cite{Kob-Andersen,Jackle,Ritort-Sollich}, and
that, for this model, the 
variation of the product of the self-diffusion constant and
the structural relaxation time is non-monotonic with temperature as a
result of the
crossover.  
The viscosity of salol seems to exhibit a 
fragile-to-strong crossover with lowering temperature
\cite{Garrahan-Chandler, Laughlin-Uhlmann}, and non-monotonic decoupling behavior
similar to what we find in our model has been observed for the
product of the shear viscosity and dielectric relaxation time in salol
\cite{Chang-Sillescu}.

\begin{figure}
  \epsfig{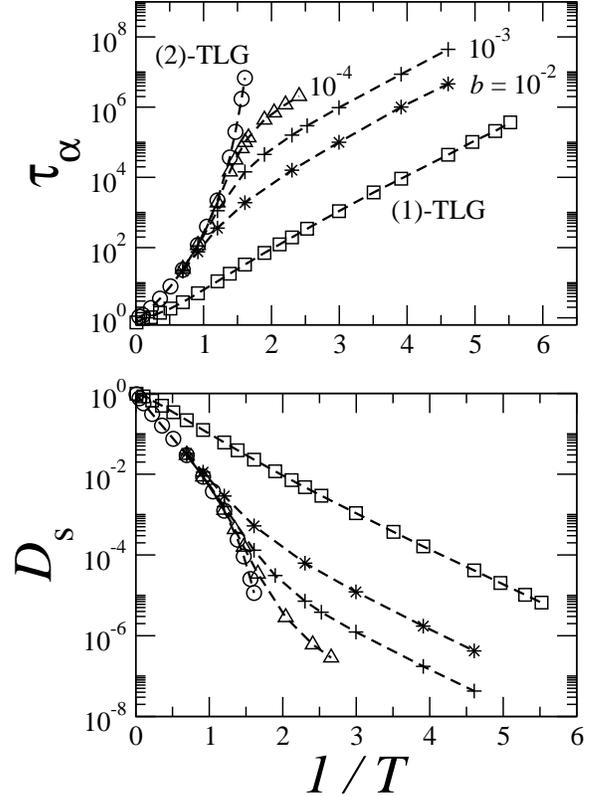}
\caption{Simulated structural relaxation time, $\tau_{\alpha}$ (top), and
 self-diffusion constant, $D_{\mbox{s}}$ (bottom), as a function of 
inverse temperature at various values of the asymmetry parameter,
 $b$: $b = 0$ (circles), $b = 10^{-2}$ (stars), $b =
 10^{-3}$ (pluses),  $b = 10^{-2}$ (triangles) and $b = 1$ (squares).  
The same labelling convention is used for all figures.  Dashed lines
 are guides to the eye and, unless otherwise indicated, statistical
 uncertainties are the size of the symbols used.}
\label{ta}
\end{figure}

Our interpolated model is based upon the kinetically constrained triangular lattice
gas (TLG) models introduced by J\"ackle and Kr\"onig \cite{Jackle}.
These two-dimensional models are variants of lattice models proposed
by Kob and Andersen \cite{Kob-Andersen}.  Each site of the triangular
lattice has six nearest neighbor sites and can hold at most one
particle.  A particle at site ${\bf i}$ is allowed to move to a
nearest neighbor site, ${\bf i}'$, if (i) ${\bf i}'$ is not occupied
and (ii) the two mutual nearest neighbor sites of ${\bf i}$ and ${\bf
i}'$ are also empty.  These rules coincide with a physical
interpretation of steric constraints on the movement of hard core
particles in a dense fluid \cite{Jackle}.  We call the model with
these rules the (2)-TLG because both mutual nearest neighbors need to
be empty in order to facilitate movement.  The model where the constraints are
more relaxed we call the (1)-TLG: movement is allowed as long as 
either of the mutual nearest neighbors is empty.
As with other kinetically constrained lattice gas models, the TLG has
no static interactions between particles other than those that
prohibit multiple occupancy of a single lattice site.  These models exhibit all the salient
features of glassformers (see, for example, \cite{Ritort-Sollich,
  Toninelli-et-al, Pan-et-al, Jackle} such as dynamical slowdown, heterogeneity and
lengthscale growth.  Moreover, the
(2)-TLG is a fragile glassformer and the (1)-TLG is a strong
glassformer \cite{Pan-et-al}.  

To study crossover behavior, we introduce a model which interpolates between
the (2)-TLG and the (1)-TLG.  A similar model which interpolates 
between the symmetrically
and asymmetrically constrained Ising chains (i.e., the
Fredrickson-Andersen \cite{Fredrickson-Andersen} and East
\cite{Jackle-Eisinger} models) predicts
crossover behavior of relaxation times \cite{Buhot-Garrahan}.  For our studies, we
introduce a parameter $b$, ranging from zero to one, which
controls the dynamics in the following way.  
When $b = 0$, the only allowed move set is that of the (2)-TLG.  When
$b>0$, a move from the (1)-TLG move set is allowed with probability $b$.
Finally, when $b=1$, the model is the (1)-TLG.  Results for the extreme cases
of $b=0$ and $b=1$ have been presented elsewhere \cite{Pan-et-al}.  In the present study, we
perform additional simulations at three intermediate values of $b$ and over
several densities, $\rho$.  At each $(\rho, b)$ state point, 32 to 64
independent trajectories of length 10-100 times the structural relaxation
time, $\tau_{\alpha}$, were collected and analyzed.  For these
simulations, $\tau_{\alpha}$ is the time for
the self-intermediate scattering function at wavevector $\pi$
to 
reach $1/e$ of its initial value.  We define a
reduced temperature, $T$, as $-\ln(1-\rho)\equiv 1/T$.  
For additional computational details, see \cite{Pan-et-al}.

The structural relaxation time in the fragile (2)-TLG scales as a double
exponential in inverse temperature and the strong (1)-TLG scales as a single
exponential in inverse temperature \cite{Toninelli-et-al, Pan-et-al}.  
We therefore anticipate the following behavior
for the crossover model.  For a value of $b$ between 0 and 1, we expect the
system to relax as if it were the fragile TLG at high temperatures.  This is because the time
needed for relaxation via the fragile TLG dynamics is faster than waiting a
time proportional to $1/b$ to relax via strong dynamics, despite the fact
that strong relaxation is faster.  At some point as temperature is
lowered, the time required to wait for a strong move becomes comparable to
the fragile relaxation time and the system crosses over to strong behavaior.
This behavior is clearly seen in FIG.\ \ref{ta}.  For every value of $b$
between 0 and 1, the structural relaxation time initially follows the (2)-TLG 
curve at high temperatures and then, at some lower temperature, $T_{\mbox{x}}$, it
crosses over to (1)-TLG behavior.  As we would expect, $T_{\mbox{x}}$
increases, as $b$ increases.

We now turn to relaxation at larger lengthscales where a similar
crossover behavior is observed in the self-diffusion constant,
$D_{\mbox{s}}$.  See FIG.\ \ref{ta}.  In a dynamically homogeneous system, one would expect the
product $D_{\mbox{s}}\tau_{\alpha}$ to be a constant independent of
temperature.  That is, one expects mean field relations such as the
Stokes-Einstein (SE) relation to be valid.  In dynamically heterogeneous
systems such as glasses, these mean field relations are dramatically
violated: relaxation at long lengthscales decouples from relaxation
at shorter lengthscales \cite{Swallen-et-al, Ediger, Jung-et-al,
  Chang-Sillescu}.  
In fragile glassformers, $D_{\mbox{s}}\tau_{\alpha}$
has been found to increase several orders of magnitude as temperature
 is lowered below some onset temperature, $T_o$, where the dynamics
starts becoming heterogeneous \cite{Swallen-et-al, Chang-Sillescu}.  
The same is true for the (2)-TLG where
the SE violation is particularly precipitous: $D_{\mbox{s}} \sim
\tau_{\alpha}^{-\xi}$ below $T_{\mbox{o}}$, where $\xi \approx 0.58$
\cite{Pan-et-al}.  
Strong glassformers, on the other
hand, are anticipated to have only a weak SE violation in 2 and 3
dimensions \cite{Jung-et-al}.  This is again true for the (1)-TLG
where, depending on the definition of $\tau_{\alpha}$, $\xi$ is
between 0.88 and 1 \cite{Pan-et-al}.

\begin{figure}
\epsfig{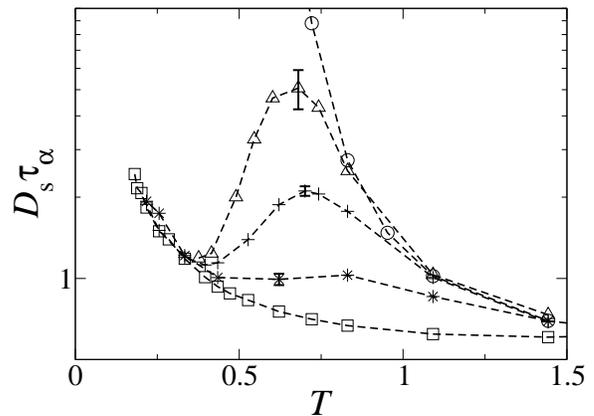}
\caption{The product of the self-diffusion constant, $D_{\mbox{s}}$, and structural
  relaxation time, $\tau_{\alpha}$, as a function of temperature at
  various values of the asymmetry parameter, $b$.  Symbols as in FIG.\
  \ref{ta}.}
\label{decoup}
\end{figure}

Therefore, for the interpolated crossover model, as $T$ is lowered, 
$D_{\mbox{s}}\tau_{\alpha}$ may vary non-monotonically.  Indeed, this 
behavior is seen in FIG.\ \ref{decoup}.  Moreover, the smaller the value
of $b$, the larger the extremum.  Since $b$ varies in the same way as
$T_{\mbox{x}}$, this implies that systems with lower
crossover temperatures will exhibit a larger decoupling extremum
 during a fragile-to-strong transition.  Of course, this trend is only strictly true if
 strong glassformers did not themselves violate SE, which is 
 approximately true for strong glassformers in 2 and 3 dimensions.
The analysis of Ref.\ \cite{Garrahan-Chandler} suggests that salol has a 
fragile to strong crossover at $T_x \approx 1.1$ $T_{\mbox{g}}$.  It is at this 
temperature where salol also shows non-monotonic decoupling \cite{Chang-Sillescu}. 
In the models we study here, we have the same coincidence of 
fragile-to-strong crossover and non-monotonic decoupling.

   \begin{figure}[b]
\epsfig{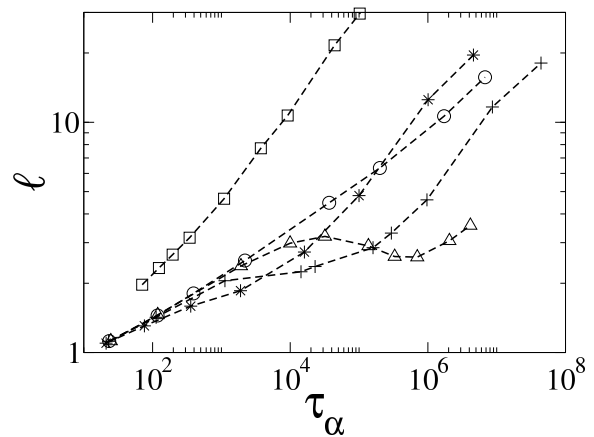}
\caption{Growth of dynamical heterogeneity length as a function of
  structural relaxation time for different values of $b$.  
  Symbols as in FIG.\ \ref{ta}.}
\label{length}
\end{figure}

Finally, we present results on the growth of dynamic heterogeneity
lengthscales during a fragile-to-strong transition.  We can study
dynamical lengthscales by
observing a trajectory over a time $\Delta t$ and considering 
the binary field $n_{\bf r}(t; \Delta t)$ = 
$p_{\bf r}(t)[1-p_{\bf r}(t+\Delta t)]$.  Here, $p_{\bf r}(t)$ is 1 if
there is a particle at lattice site ${\bf r}$ at time $t$ and 0
otherwise.  The field, $n_{\bf r}(t; \Delta t)$, gives
a signal only when there is particle motion at ${\bf r}$ over the
range of time between $t$ and $t+\Delta t$.  We then extract a
dynamical lengthscale, $\ell(\Delta t)$, as the inverse of the
first moment of the structure factors of the mobile particles
\cite{Pan-et-al, Glotzer,Garrahan-Chandler,Lacevic-et-al,Berthier}.  
These structure factors are four point functions: they measure a correlation function which depends
on two points in time, $t$ and $t + \Delta t$, and two points in
space, ${\bf r}$ and ${\bf r'}$.  The growth of the dynamic
heterogeneity lengthscale with structural relaxation time also
exhibits crossover behavior.  At some crossover lengthscale, $\ell_{\mbox{x}}$,
the growth changes from fragile growth to strong growth.  This
behavior is consistent with \cite{Garrahan-Chandler}, where it is
argued that generally the relaxation of all fragile systems will obey strong
dynamics at large enough lengthscales.  We predict, at least for
systems exhibiting entropic crossovers \cite{entropy} like the one studied here, 
that lower crossover temperatures, $T_{\mbox{x}}$, will give rise to 
larger decoupling extremum as temperature is lowered through the
crossover region.
  
\acknowledgments 
We thank Rob Jack for an important discussion.  
This work was supported by the US National Science Foundation, 
by the US Department of Energy grant no.\ DE-FE-FG03-87ER13793, by
EPSRC 
grants no.\ GR/R83712/01 and GR/S54074/01, and University of
Nottingham grant no.\ FEF 3024.
A.C.P. is an NSF graduate research fellow.


\begin{thebibliography}{99}
 
\bibitem{Ediger-et-al} M.D. Ediger, C.A. Angell and S.R. Nagel,
J. Phys. Chem. {\bf 100}, 13200 (1996).

\bibitem{Angell} C.A. Angell, Science {\bf 267}, 1924 (1995).

\bibitem{Debenedetti-Stillinger} P.G. Debenedetti and F.H. Stillinger,
Nature {\bf 410}, 259 (2001).
 
\bibitem{Ito-et-al} K. Ito, C.T. Moynihan and C.A. Angell, Science {\bf 298}, 492 (1999).

\bibitem{Buhot-Garrahan} A. Buhot and J.P. Garrahan, Phys. Rev. E {\bf 64}, 021505
(2001).

\bibitem{Garrahan-Chandler} J.P. Garrahan and D. Chandler, Proc. Natl. Acad. Sci. USA
{\bf 100}, 9710 (2003).


\bibitem{Ritort-Sollich} F. Ritort and P. Sollich, Adv. Phys. {\bf
52}, 219 (2003).

\bibitem{Jackle} J. J\"ackle and A. Kr\"onig, J. Phys. Condens. Matter
{\bf 6}, 7633 (1994); 7655 (1994).

\bibitem{Kob-Andersen} W. Kob and H.C. Andersen, Phys. Rev. E {\bf
48}, 4364 (1993).


\bibitem{Laughlin-Uhlmann} W.T. Laughlin and D.R. Uhlmann,
  J. Phys. Chem. {\bf 76}, 2317 (1972).

\bibitem{Chang-Sillescu} I. Chang and H. Sillescu, J. Phys. Chem. B
{\bf 101}, 8794 (1997).

\bibitem{Pan-et-al} A.C. Pan, J.P. Garrahan and D. Chandler,
cond-mat/0410525.

\bibitem{Fredrickson-Andersen} G. H. Fredrickson and H. C. Andersen, 
Phys. Rev. Lett. \textbf{53}, 1244 (1984).

\bibitem{Jackle-Eisinger} J. J\"ackle and S. Eisinger, 
Z. Phys. B 84\textbf{84}, 115 (1991).

\bibitem{Toninelli-et-al} C. Toninelli, G. Biroli and D.S. Fisher,
Phys. Rev. Lett. {\bf 92}, 185504 (2004); C. Toninelli and G. Biroli,
cond-mat/0402314.

\bibitem{Jung-et-al} Y.J. Jung, J.P. Garrahan and D. Chandler,
Phys. Rev. E {\bf 69}, 061205 (2004).
 
\bibitem{Ediger} M.D. Ediger, Annu. Rev. Phys. Chem.  {\bf 51}, 99
  (2000).

\bibitem{Swallen-et-al} S.F. Swallen, P.A. Bonvallet, R.J. McMahon and
M.D. Ediger, Phys. Rev. Lett. {\bf 90}, 015901 (2003).
 
\bibitem{Jack-et-al} R. L. Jack, J. P. Garrahan and L. Berthier,
 unpublished.

\bibitem{Glotzer} S.C. Glotzer, J. Non-Cryst. Solids, {\bf 274}, 342
(2000).
 
\bibitem{Lacevic-et-al} N. Lacevic, F.W. Starr, T.B. Schr{\o}der and
S.C. Glotzer, J. Chem. Phys. {\bf 119}, 7372 (2003).

\bibitem{Berthier} L. Berthier, Phys. Rev. E \textbf{69}, 020201(R)
(2004).

\bibitem{entropy} Entropic crossovers occur in systems where the
  directional persistence probability of particle motion is high and
  primarily caused by
 entropy (see Ref. \cite{Garrahan-Chandler}).

\end{thebibliography}
\end{document}